\documentstyle[12pt,epsf]{article}

\def\rmp{{\rm p}}
\def\rmk{{\rm k}}
\def\bfp{{\bf p}}
\def\bfk{{\bf k}}
\def\bfx{{\bf x}}
\def\GeV{Ge$\!$V}
\begin{document}


\title{Threshold Top Quark Production in $e^+e^-$ Annihilation:
       Rescattering Corrections\footnote{Talk presented at the XXXVI Cracow
         School of Theoretical Physis, Zakopane, June 1--10, 1996}}
\author{M. Peter \\
     {\normalsize Institut f\"ur Theoretische Teilchenphysik,} \\
     {\normalsize Universit\"at Karlsruhe,} \\
     {\normalsize D-76128 Karlsruhe, Germany}}
\date{\today}
\maketitle
\thispagestyle{empty}
\vspace{-4.0truein}
\begin{flushright}
{\bf TTP 96-36\footnote{
    The complete paper, including figures, is
    also available via anonymous ftp at
    ttpux2.physik.uni-karlsruhe.de (129.13.102.139) as
    /ttp96-36/ttp96-36.ps,
    or via www at
    http://www-ttp.physik.uni-karlsruhe.de/cgi-bin/preprints}
}\\
{\bf August 1996}\\
{\bf hep-ph/9608436}
\end{flushright}
\vspace{3.0truein}
\begin{abstract}
  Top quark pair production close to threshold at a next linear
  collider is discussed, with special focus on the influence of
  rescattering corrections. Numerical results are shown for the
  differential cross-section, i.e.\ the momentum distribution and the
  forward backward asymmetry, and for moments of the angular
  distribution of leptons arising from the semileptonic top quark
  decay.
\end{abstract}
%
%
\section{Introduction}

The top quark certainly is one of the most interesting of the presently known
elementary particles, partly because it is the one we know least of, but also
because its large mass implies that we can use it for a variety of
unprecedented QCD studies. A future 500 \GeV $e^+e^-$ collider will provide a
clean laboratory where we can investigate the top quark's properties to a high
precision by performing a detailed analysis of its production close to
threshold. It is thus not surprising to find a long series of papers concerned
with this subject, i.e.\ \cite{FaKh}--\cite{HJKP96}, leading to the following
conclusions:
\begin{enumerate}
\item The excitation curve $\sigma(E)$, i.e.\ the total cross-section for 
  $t\bar t$-production, allows for a precise measurement of $\alpha_{\rm s}$
  and the top quark mass. This is in analogy to the charmonium and
  bottomonium systems with the difference that perturbative QCD is far more
  reliable in the $t\bar t$ system because of the considerably higher scale
  and, as will be explained below, the large top quark width.
\item The momentum distribution $d\sigma/d|\bfp|$ and the forward-backward
  asymmetry will provide a second, independent determination of $\alpha_{\rm 
  s}$ and also measure the top quark width $\Gamma_t$ and consequently give
  us a handle on $V_{\rm tb}$ or physics beyond the standard model. 
\item The size of the top quark polarization, which is predicted to amount
  to $\approx -40\%$ even for unpolarized beams, is governed by the top
  quark's couplings to the $Z$-boson and the photon. Hence its measurement
  will give us information about these presently unknown parameters.
\end{enumerate}
We will also gain some information about the QCD potential, which
determines the dynamics of the nonrelativistic ``toponium'' system.

When compared with charmonium and bottomonium there seems to be a big
problem connected to the top quark, caused by its enormous width of about
1.5 \GeV: the heavy quarks will decay before resonances can build up and thus
we won't see narrow peaks in $\sigma(E)$.  Furthermore, the width cannot be
neglected in calculations.

But as Fadin and Khoze pointed out several years ago \cite{FaKh}, this obstacle
in fact turns out to be an advantage, because it implies that top decays before
hadronization effects can spoil the purely perturbative behaviour.  The width
acts as an infrared cutoff and protects us from the poorly known physics at
large distances. In addition, the study of momentum distributions and top
polarization is only made possible by the dominance of the single quark decay.
$J/\psi$ and $\Upsilon$ for example mainly decay via $q\bar q$ annihilation and
thus all information about momentum distributions is lost.

In the following section I will briefly describe the calculation of the
cross-section for $t\bar t$ production including terms of the order $\beta$.
These results are however inconsistent without the inclusion
of ${\cal O}(\alpha_{\rm s})$ corrections, which I will present in section 3
for the differential cross-section and in section 4 for the top quark
polarization. I will not dwell too much on the technical parts of the
calculation. The reader who is interested in more details is advised to read
\cite{HJKP96}.
%
%
\section{``Born-level'' results for $t\bar t$ production}

\begin{figure}[htb]
   \begin{flushleft} \epsfxsize 5cm \mbox{\epsfbox[0 30 195 100]{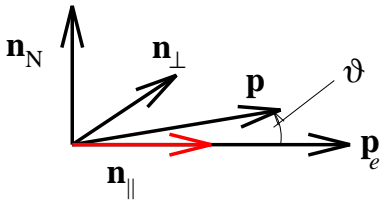}} 
   \hfill\parbox{8cm}{\caption{\label{fig1}Orientation of the reference
       frame.}}\end{flushleft}
\end{figure}

Close to threshold it is convenient to employ a nonrelativistic
approximation and use the corresponding variables: the top quark
three-momentum in the laboratory frame $\bfp$, with $\rmp=|\bfp|$ and
$\beta=\rmp/m_t$, and the energy measured from nominal threshold $E=\sqrt
s-2m_t={\cal O}(\beta^2)$.  We will also need the angle $\vartheta$ between
the top and electron beam directions and the three unit vectors ${\bf
  n}_{_\parallel}$, ${\bf n}_{_\perp}$ and ${\bf n}_{\rm_N}$ which are
defined in Fig.~\ref{fig1}.

In the kinematic regime $\beta\ll1$ one usually encounters the problem that
Coulomb gluon exchange is not suppressed because $\alpha_{\rm s}/\beta$ is
not a small parameter and thus all terms of the form $(\alpha_{\rm
  s}/\beta)^n$ have to be summed up. This is usually solved by using the
Schr\"odinger equation to calculate the wave-functions for the bound states
and expressing the observables in terms of the latter. In the $t\bar t$ system
there is however the additional complication already mentioned: the large
width which cannot be neglected.  As a consequence there will be no isolated
resonances in the cross-section and thus the usage of wave-functions is very
inconvenient, because lots of them would have to be summed, including
interferences.

The idea of Fadin and Khoze \cite{FaKh} to circumvent this was to use
the Green function instead, which by construction is a sum over all
wave-functions:
$$ G({\bf r},{\bf r}^\prime;E) = \sum_{\rm states~n}\frac{\psi_n^*({\bf r})
   \psi_n({\bf r}^\prime)}{E-E_n+i\varepsilon}.
$$

\begin{figure}
   \begin{center} \epsfxsize 10cm \mbox{\epsfbox{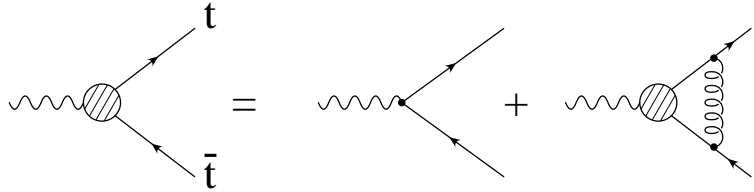}} \end{center}
  \caption{\label{lseq}Graphical representation of the Lippmann-Schwinger
    equations.}
\end{figure}

A method to see this function appear in a diagrammatical approach in
momentum space is as follows\footnote{This method has first been used in
  \cite{SP91}}: one can express the fact that all gluon exchanges have to be
summed (in ladder approximation) in the form of self-consistency equations for
the vertex functions $\Gamma_{\cal C}$, with ${\cal C}=\gamma^\mu$ or
$\gamma^\mu\gamma^5$ in the case of interest for $e^+e^-$ annihilation,
\begin{equation}
  \Gamma_{\cal C} = {\cal C} + \int\!{d^4k \over (2 \pi)^4}
  \left( -{4 \over 3} 4 \pi \alpha_{\rm s} \right)
    D_{\mu \nu}(p-k) \gamma^\mu S_{\rm F}(k+{q \over 2})
    \Gamma_{\cal C}(k,q) S_{\rm F}(k-{q\over2}) \gamma^\nu ,
\end{equation}
which are also shown in Fig.\ \ref{lseq}. In the nonrelativistic approximation,
neglecting terms of the order $\beta^2$,
they can essentially be reduced to the integral equations\footnote{See
\cite{HJKP96} for details of the derivation}
\begin{eqnarray}
{\cal K}_{\rm S} (\rmp,E) &=& 1 + \int {d^3 k \over (2 \pi)^3}
        V(\bfp - \bfk) {{\cal K}_{\rm S} (\rmk,E) \over E -
        {\bfk^2 \over m_t} + i \Gamma_t}  \\
{\cal K}_{\rm P} (\rmp,E) &=& 1 + \int {d^3 k \over (2 \pi)^3}
        {\bfp\cdot \bfk \over \bfp^2}
        V(\bfp - \bfk) {{\cal K}_{\rm P} (\rmk,E) \over E -
        {\bfk^2 \over m_t} + i \Gamma_t}.
\end{eqnarray}
The functions ${\cal K}$ in turn are tightly connected to the Green function
${\cal G}$ which solves the Lippmann-Schwinger equation
\begin{equation}\label{LS}
\bigg[E - {\bfp^2 \over m_t} + i \Gamma_t \bigg]
        {\cal G} (\bfp,\bfx,E) = e^{i \bfp \cdot \bfx} +
        \int {d^3 k \over (2 \pi)^3} V(\bfp - \bfk)
        {\cal G} (\bfk,\bfx,E) :
\end{equation}
expanding ${\cal G}$ in a Taylor series around ${\bf x}=0$ we find
$$ {\cal G}(\bfp,\bfx,E) = G(\rmp,E) + \bfx \cdot i \bfp \, 
   F(\rmp,E) + \ldots
$$
with
\begin{equation}
  G(\rmp,E) = G_0(\rmp,E) {\cal K}_{\rm S} (\rmp,E) , \qquad
  F(\rmp,E) = G_0(\rmp,E) {\cal K}_{\rm P} (\rmp,E)
\end{equation}
and
\begin{equation}\label{free}
G_0(\rmp,E) = {1\over E - {\bfp^2 \over m_t} + i \Gamma_t}.
\end{equation}
$G$ and $F$ thus are the $S$- and $P$-wave Green functions, respectively. They
form the nontrivial dynamical ingredients for the production cross-section
which is otherwise completely determined by electroweak coupling constants.
For a realistic potential they have to be determined numerically, and it turns
out convenient to perform this calculation directly in momentum space using a
method introduced in \cite{JKT92}.

We will employ the following conventions for the fermion couplings
\begin{equation}
  v_f = 2 I^3_f - 4 q_f \sin^2\theta_{\rm W} , \qquad  a_f = 2 I^3_f
\end{equation}
and the useful abbreviations
\begin{equation}\begin{array}{ll}
  a_1 = (q_eq_t+v_ev_td_Z)^2 + (a_ev_td_Z)^2 &
  a_2 = 2a_ev_td_Z(q_eq_t+v_ev_td_Z) \\
  a_3 = 2a_ea_td_Z(q_eq_t\!+\!2v_ev_td_Z) &
  a_4 = 2a_td_Z(q_eq_tv_e\!+\!(v_e^2\!+\!a_e^2)v_td_Z) \\
  d_Z = \frac{s}{(s-m_Z^2)}\frac{1}{16\sin^2\Theta_W\cos^2\Theta_W}.
\end{array}\end{equation}
$P_\pm$ denotes the longitudinal positron/electron polarization and the
variable $\chi$ is defined as
\begin{equation}\label{chi}
\chi={P_+-P_-\over1-P_+P_-}.
\end{equation}
With these parameters the differential cross-section for the
production of a $t\bar t$ pair with a top quark of three-momentum
$\bfp$ and spin projection $1/2$ on direction {\bf n} can be written
in the form
\begin{eqnarray}
 \frac{d^3\sigma}{dp^3} &=& \frac{1}{4\pi\rmp^2}\frac{d\sigma}{d\rmp}\Big(
   1+2{\cal A}_{\rm _{FB}}(\rmp,E)\cos\vartheta\Big)\frac{1+{\bf P}\cdot
     {\bf n}}{2} \label{dsigma}\\
 \frac{d\sigma}{d\rmp} &=& \frac{3\alpha^2}{m_t^4}\Big(1-P_+P_-\Big)\Big(
   a_1+\chi a_2\Big)\cdot\Gamma_t\cdot|\bfp\cdot G(\rmp,E)|^2 .
\end{eqnarray}

${\cal A}_{\rm_{FB}}$ denotes the (momentum dependent) forward-backward
asymmetry and {\bf P} can be interpreted as top quark polarization vector.
These quantities are given by
\begin{eqnarray}
 {\cal A}_{\rm_{FB}}(\rmp,E) &=& -C_{\rm_N}(\chi)\, \varphi_{\rm _R}(\rmp,E) \\
 {\rm P}_{_\parallel}(\rmp,E) &=& C_{_\parallel}^0\,+\,C_{_\parallel}^1
    \cdot\varphi_{\rm _R}(\rmp,E)\cos\vartheta\\
 {\rm P}_\perp(\rmp,E)  &=& C_\perp\cdot\varphi_{\rm _R}(\rmp,E)
    \sin\vartheta \\
 {\rm P}_{\rm _N}(\rmp,E)  &=& C_{\rm _N}\cdot\varphi_{\rm _I}
 (\rmp,E)\sin\vartheta
\end{eqnarray}
with ${\rm P}_{_\parallel,\perp,\rm _N}={\bf P}\cdot{\bf n}_{_\parallel,\perp,
\rm _N}$,
\begin{eqnarray}
  C_{_\parallel}^0(\chi) & = & -\frac{a_2+\chi a_1}{a_1+\chi a_2} \\
  C_{_\parallel}^1(\chi) & = & \big(1-\chi^2\big)\frac{a_2a_3-a_1a_4}
                            {(a_1+\chi a_2)^2} \\
  C_\perp(\chi) & = & -\frac{1}{2}\frac{a_4+\chi a_3}{a_1+\chi a_2} \\
  C_{\rm _N}(\chi) & = & \frac{1}{2}\frac{a_3+\chi a_4}{a_1+\chi a_2},
\end{eqnarray}
\begin{equation}\label{phi}
 \varphi(\rmp,E) = \frac{\rmp}{m_t}\cdot\frac{F^*(\rmp,E)}{G^*(\rmp,E)}
\end{equation}
and $\varphi_{\rm _R} = \mbox{Re}\,\varphi$, $\varphi_{\rm _I} = \mbox{Im}\,
\varphi$. They have first been derived for polarized beams in \cite{HJKT95}, 
apart from P$_{\rm _N}$ which has been supplemented in \cite{HJKP96}. The
forward-backward asymmetry has been calculated first, for unpolarized beams,
in \cite{MS93}.

\begin{figure}[ht]
 \begin{center}
  \epsfxsize=12cm \mbox{\epsfbox[40 290 535 525]{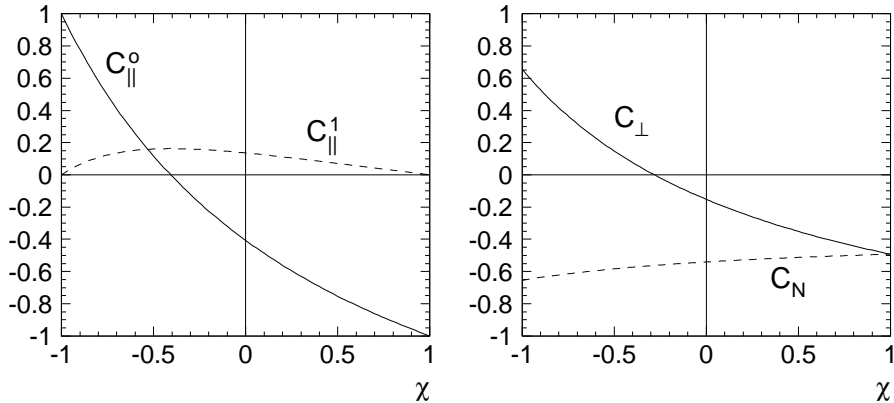}}
  \caption[]{\label{Cx}\sloppy
        The coefficient functions $C_x$ for $\sqrt{s}/2 = m_t=180$ GeV.}
 \end{center}
\end{figure}

The coefficient functions $C_x$ are displayed in figure \ref{Cx}. It is not
difficult to see that for small $\beta$ the top quark spin is mainly
aligned with the electron beam direction, apart from corrections
of order $\beta$ that depend on the production angle $\vartheta$.

However, the size of these ${\cal O}(\beta)$ corrections is not really given
by the expressions above, although they are correct in a formal sense.
Close to threshold $\beta \sim \alpha_{\rm s}$ and consequently
${\cal O}(\alpha_{\rm s})$ terms might be as large, or put another way,
they should be included in the ${\cal O}(\beta)$ analysis.

Some of these QCD corrections are not too difficult to implement
because they are universal. The $\gamma t\bar t$ and $Zt\bar t$ vertex
corrections caused by the exchange of transverse gluons for example
can be incorporated to the required order by simply multiplying with
the following factors \cite{KZ88}:
\begin{eqnarray}
  G(\rmp,E) & \longrightarrow & \Big(1-\frac{8\alpha_{\rm s}}{3\pi}\Big)
     G(\rmp,E) \\
  F(\rmp,E) & \longrightarrow & \Big(1-\frac{4\alpha_{\rm s}}{3\pi}\Big)
     F(\rmp,E)
\end{eqnarray}
resulting in a corresponding modification in the definition of
$\varphi(\rmp,E)$. Corrections to the individual top decays can be accounted
for by using the 1-loop result for $\Gamma_t$ \cite{JK89}.

\begin{figure}[h]\begin{center}
\begin{tabular}{cc}
  \epsfxsize 63mm \mbox{\epsffile[25 570 200 650]{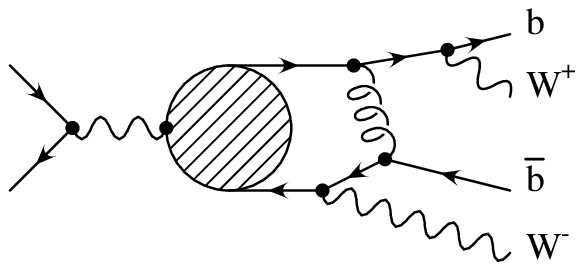}} &
  \epsfxsize 63mm \mbox{\epsffile[25 570 200 650]{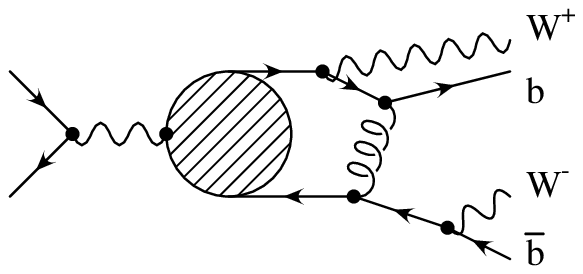}} \\
  a) & b)
\end{tabular}\end{center}
\caption{\label{fig:resc}Rescattering diagrams.}
\end{figure}

There is however a type of corrections which is more difficult because it
connects the different stages of the production and subsequent decay
process. I will call them ``rescattering corrections'', the corresponding
diagrams are depicted in figure \ref{fig:resc} (in principle there is a
third diagram with gluon exchange between the two $b$-quarks, which can
however be neglected to the required order \cite{HJKP96}).

Fortunately, their contribution to the total cross-section vanishes \cite{
  MS93,MY94}, but as has first been shown in \cite{MS93}, the contribution to
the forward-backward asymmetry does not (although it was found to be small in a
numerical study \cite{FMS94}). And there is no a priori argument why they
should be irrelevant for the polarization vector, especially for those
components which start at order $\beta$, i.e\ P$_\perp$ and P$_{\rm _N}$.  For
this reason I will discuss rescattering corrections in the following two
sections.
%
%
\section{Corrections to the differential cross-section}

By making use of the nonrelativistic approximation, the calculation of the
rescattering corrections to the differential cross-section can essentially be
performed along the same line as that of the leading order result. Therefore
I will skip that part and only discuss the result. The reader who is
interested in the technical aspects should consult \cite{HJKP96} or
\cite{MS93}.

Defining the two convolution functions
\begin{eqnarray}
   \psi_1(\rmp,E) &=& 2\,\mbox{Im}\int\!\frac{d^3k}{(2\pi)^3}V(|\bfk-\bfp|)
        \frac{G(\rmk,E)}{G(\rmp,E)}\frac{\arctan{\frac{|\bfk-\bfp|}
        {\Gamma_t}}}{|\bfk-\bfp|} \label{psi1}\\
   \psi_2(\rmp,E) &=& 2\int\!\frac{d^3k}{(2\pi)^3}V(|\bfk-\bfp|)\frac{\bfp
       \cdot({\bf k-p})}{\rmp|{\bf k-p}|^2}\frac{G(\rmk,E)}{G(\rmp,E)}
        \nonumber \\ && \times
        \Big(1-\frac{\Gamma_t}{|\bfk-\bfp|}\arctan{\frac{|\bfk-\bfp|}{\Gamma_t}
        }\Big) \label{psi2}\\
   \psi_{\rm _R}(\rmp,E) &=& \mbox{Re}\;\psi_2(\rmp,E),
\end{eqnarray}
the corrected formula for $d^3\sigma/dp^3$ reads
\begin{equation}
  \frac{d^3\sigma}{dp^3} = \frac{1}{4\pi\rmp^2}\frac{d\sigma_{\rm Born}}{
    d\rmp}\cdot\Big(1+\psi_1(\rmp,E)\Big)\cdot\Big(1+2{\cal A}^{(1)}_{
    \rm_{FB}}(\rmp,E)\cos\vartheta\Big)
\end{equation}
with
\begin{equation}
   {\cal A}^{(1)}_{\rm_{FB}}(\rmp,E) = {\cal A}_{\rm_{FB}}(\rmp,E) +
      \frac{1}{2}\frac{1-2y}{1+2y}\;C_{_\parallel}^0\cdot\psi_{\rm _R}(\rmp,E)
\end{equation}
and $y=m_W^2/m_t^2$. Rescattering thus has two effects:

\begin{figure}[t]
\begin{tabular}{cc}
  \epsfxsize 65mm \mbox{\epsffile[0 0 567 454]{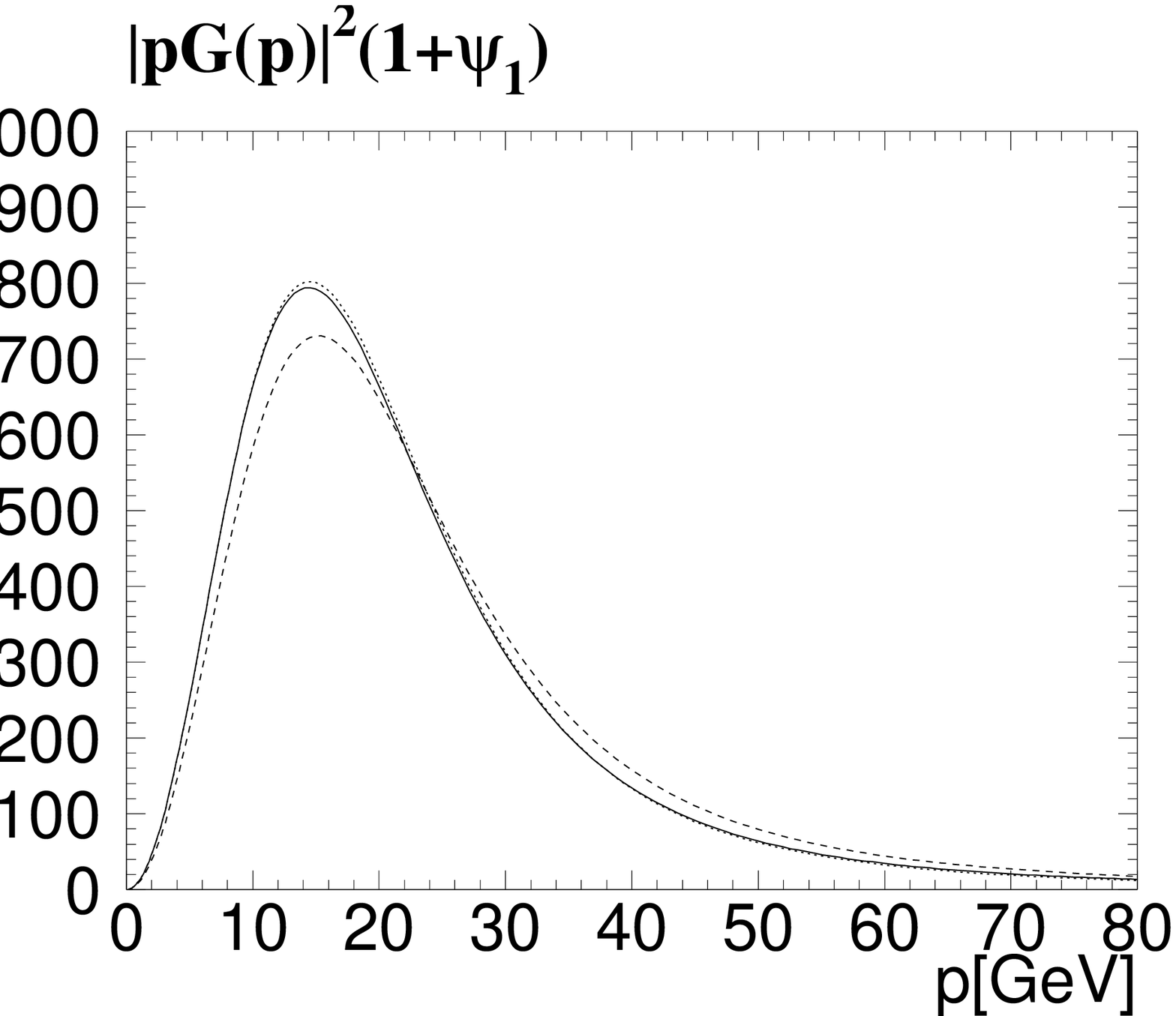}} & 
  \epsfxsize 65mm \mbox{\epsffile[0 0 567 454]{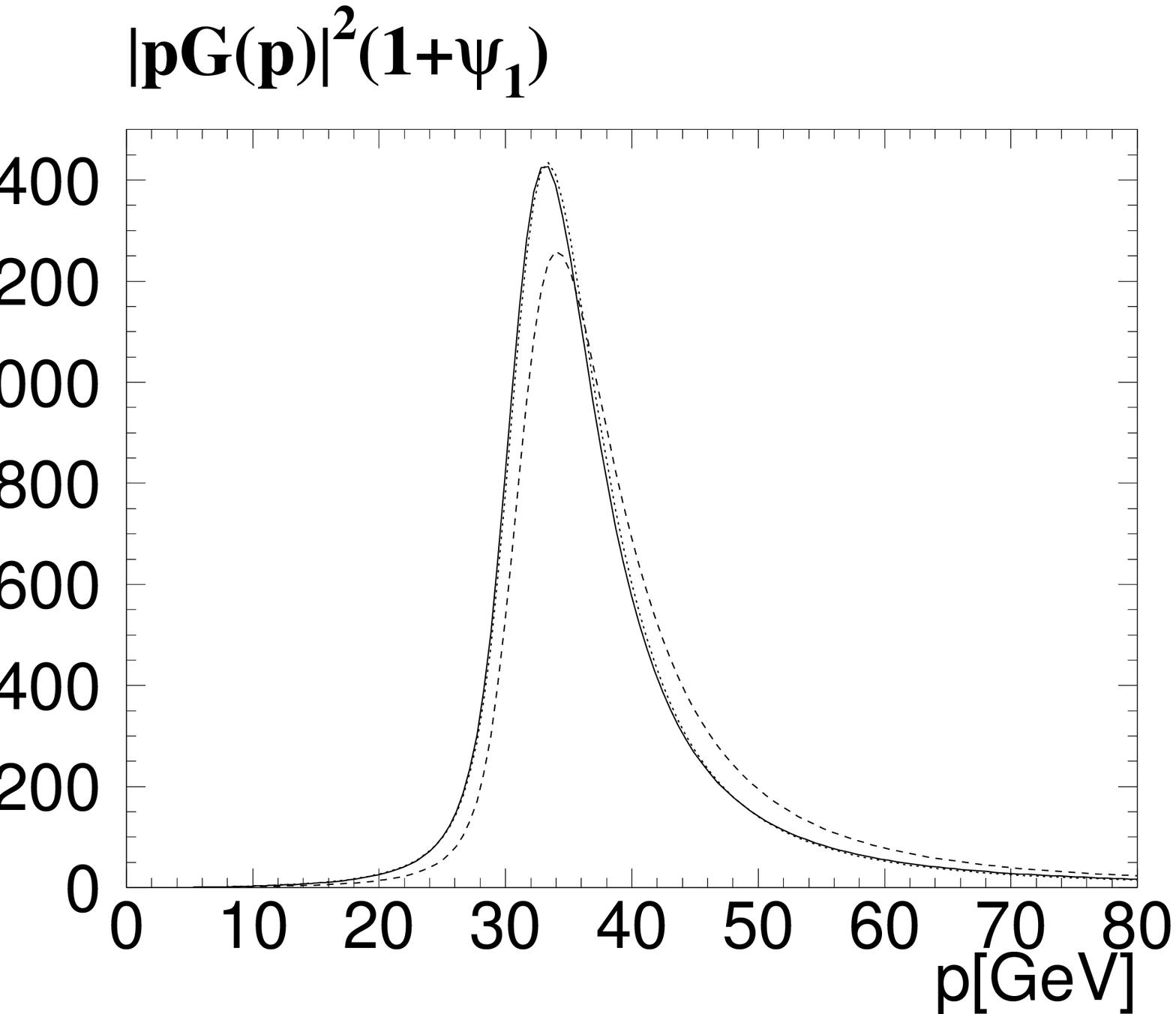}} \\
   a) E=-3\GeV & b) E=5\GeV  
\end{tabular}
\caption{Shape of the momentum distribution of the top quarks for $E=3$\GeV
  (the position of the would-be 1S resonance) and $E=5$\GeV.  Dashed line:
  no rescattering corrections included; solid line: rescattering
  contribution using the full potential included; dotted line: rescattering
  contribution included using a pure Coulomb potential with $\alpha_{\rm
    s}=0.187$.
\label{fig:dsigma}}
\end{figure}

\begin{figure}[htb]
\begin{tabular}{cc}
  \epsfxsize 65mm \mbox{\epsffile[0 60 567 760]{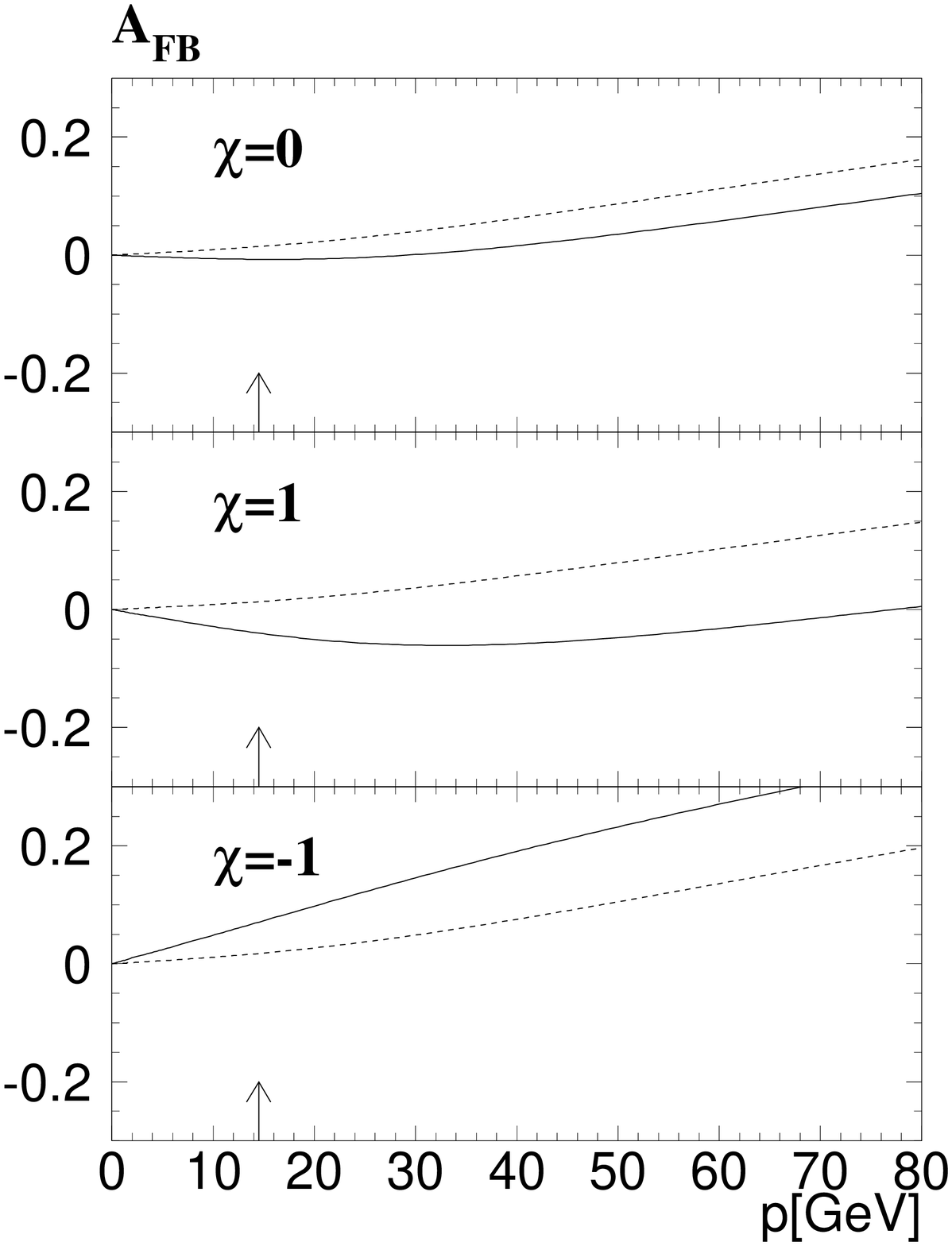}} &
  \epsfxsize 65mm \mbox{\epsffile[0 60 567 760]{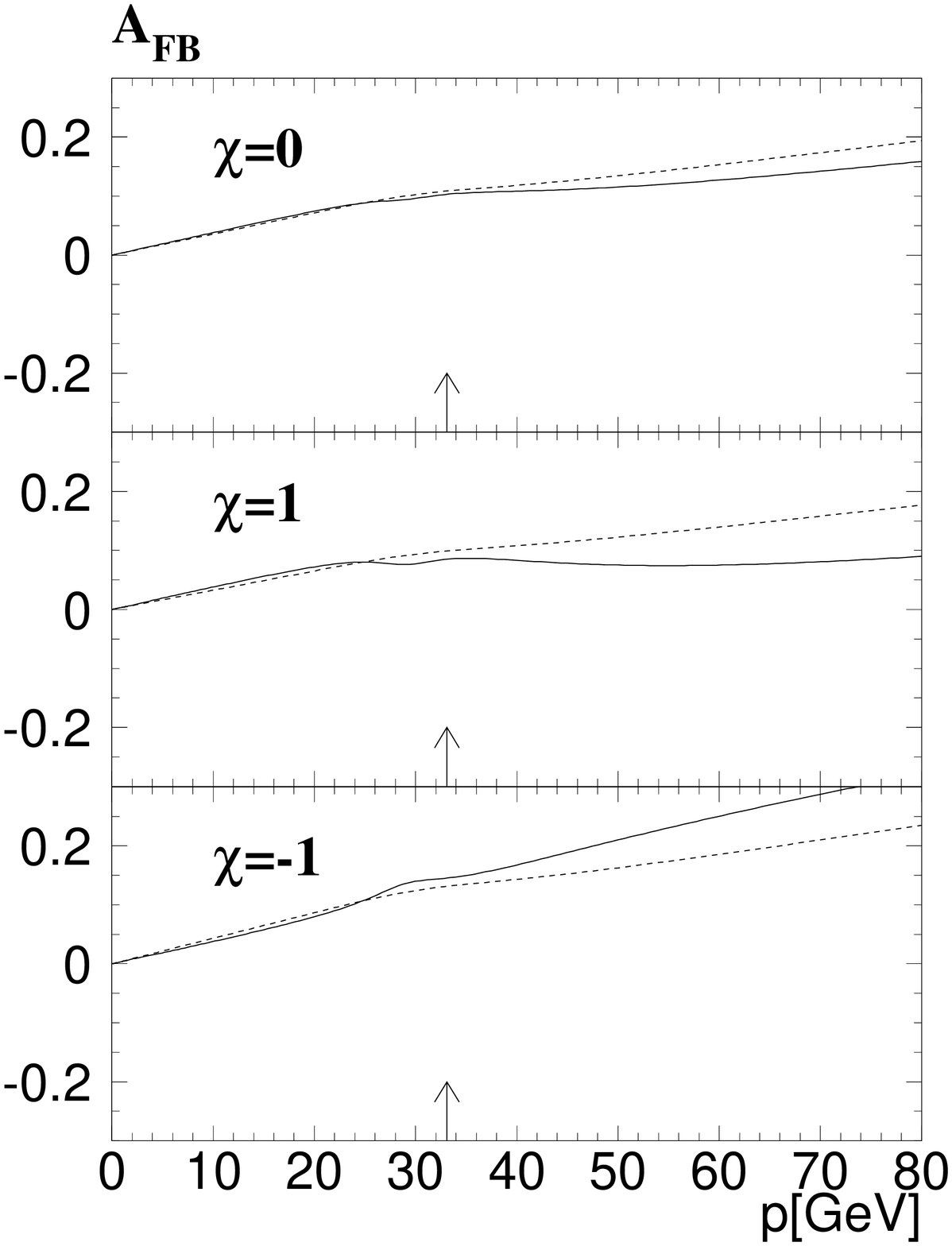}} \\
   a) E=-3\GeV & b) E=5\GeV
\end{tabular}
\caption{\label{fig:afb}Rescattering corrections to the forward-backward
asymmetry. Dashed line: $S$-$P$--wave interference contribution; solid line:
full result. The little arrows indicate the position of the peak in the
momentum distribution.}
\end{figure}

\begin{enumerate}
\item There is a shift in the momentum distribution, as visible in the factor
  $1+\psi_1(\rmp,E)$, which has a simple physical explanation: if $\bar t$
  decays before $t$, its decay product $\bar b$ attracts the top quark and in
  this way reduces the measured momentum. Figure \ref{fig:dsigma} displays the
  size of this effect. The dashed lines show the leading order result, the
  solid and (hardly visible) dotted lines are the corrected curves, with the
  correction implemented in two different ways: the solid line was calculated
  using the same potential\footnote{Its actual expression can be found in
    \cite{JT93}} in the determination of the Green function $G$ and the
  function $\psi_1$; the dotted line was calculated using a pure Coulomb
  potential in Eq.\ \ref{psi1}, with the coupling constant fixed to
  $\alpha_{\rm s}=0.187$ in order to reproduce the solid line. This was done
  to compare with \cite{MS93} where a pure Coulomb potential was used as well,
  but with $\alpha_{\rm s}=\alpha_{\rm s}(\alpha_{\rm s}(M_Z)m_t)\approx
  0.15$. Of course the scale governing rescattering corrections is of the
  order of the top momentum $\rmp\sim\alpha_{\rm s}(M_Z)m_t$, but when using a
  fixed coupling constant, its actual value is nevertheless unknown a priori.
\item There is a change in the forward-backward asymmetry which depends
  on the beam polarization, as can be seen from Fig.\ \ref{fig:afb}. The
  dashed line again shows the uncorrected, the solid line the corrected result.
  The small arrows indicate the position of the peak in the momentum
  distribution, i.e.\ the region which is most relevant. The reason why
  the corrections were found to be small in \cite{MS93} is that these
  authors only considered unpolarized beams.
\end{enumerate}

There are also a few things worth noticing:

\noindent
The role of $\Gamma_t$ as an infrared cut-off is obvious in Eq.\
  \ref{psi2}. For vanishing width the integrand would behave as 
  $$ \frac{V(|\bfk-\bfp|)}{|\bfk-\bfp|} \sim \frac{1}{|\bfk-\bfp|^3} $$
  for $\bfk\sim\bfp$, but the factor in brackets weakens this behaviour
  considerably:
  $$ \frac{V(|\bfk-\bfp|)}{|\bfk-\bfp|}\Big(1-\frac{\Gamma_t}{|\bfk-\bfp|}
    \arctan{\frac{|\bfk-\bfp|}{\Gamma_t}}\Big) \sim \frac{1}{3\Gamma_t^2}
    \frac{1}{\rmk},\qquad|\bfk-\bfp|\ll\Gamma_t.
  $$

The corrections to the total cross-section indeed vanish because
  they are proportional to the integral
  $$ \int\!\frac{d^3p}{(2\pi)^3}|G(\rmp,E)|^2\psi_1(\rmp,E)=0.$$
  This can be seen by inserting Eq.\ \ref{psi1} and interchanging the
  integration variables $\bfp$ and $\bfk$.

It is advantageous to solve the Lippmann-Schwinger equations
  in momentum space because this allows for a relatively easy numerical
  deter\-mi\-nation of $\psi_1$ and $\psi_2$. In the coordinate space approach
  additional Fourier transformations would be necessary leading to a loss
  in accuracy.

%
\section{Top quark polarization}

The calculation of the rescattering corrections to top quark polarization is
more involved than that of the spin-averaged differential cross-section. In the
Born-level calulation the quark is essentially treated as a stable particle,
the free Green function $G_0$ being the only place where its decay enters
explicitly through the usage of the complex energy $E+i\Gamma_t$. The top quark
spin is considered a well defined quantity and thus the appropriate projection
operators are used in the derivation of Eq.\ \ref{dsigma}.

This is however no longer possible in the evaluation of the diagram shown in
Fig.\ \ref{fig:resc}(b). There the top quark appears as a virtual particle
only and thus the notion of its polarization is highly questionable. To find
a way out one may ask: How will the top quark polarization be measured in a
real experiment ?

The answer to this question is known: the experimentalists will use the
angular distribution of the leptons from the decay $t\to bl\nu$, because
for a free top quark in its rest frame, this is given by
$$ \frac{d\Gamma_{t\to bl\nu}({\bf P})}{d\cos\theta}
   = \frac{1}{2}\Gamma_{t\to bl\nu}\Big(1+|{\bf P}|\cos\theta\Big) $$
where {\bf P} denotes the top quark polarization vector and $\theta$ the
angle between {\bf P} and the lepton momentum.

The generalization of this formula to $t\bar t$ production in $e^+e^-$
collisions would be
\begin{equation}\label{fact}
   \frac{d^6\sigma}{dp^3\;dE_l\;d\Omega_l} = \frac{d^3\sigma}{dp^3}\cdot
   \frac{1}{\Gamma_t}\frac{d\Gamma_{t\to bl\nu}({\bf P})}{dE_l\;d\Omega_l}
\end{equation}
which expresses the fact that first a top quark with momentum \bfp\ and
polarization {\bf P} has to be produced, which may then decay
semileptonically. 
Diagram \ref{fig:resc}(b) however destroys this factorization because
the production and decay stages are no longer separated.

Of course the left hand side of Eq.\ \ref{fact} could be calculated
nevertheless, at least in principle, however it would lead to a rather
complicated result. A method to obtain simpler expressions involving an easier
calculation is to consider moments of the lepton spectrum as defined by
\begin{equation}
  \langle nl\rangle \equiv
       \Big(\frac{d^3\sigma}{dp^3}\Big)^{-1}\int dE_ld\Omega_l
       \frac{d\sigma(e^+e^-\to bl\nu\bar bW^-)}{dp^3\;d\Omega_p\;d\Omega_l
         \;dE_l}(nl)
\end{equation}
or
\begin{equation}
  \langle\langle nl\rangle\rangle \equiv 
       \Big(\frac{d\sigma}{d\rmp}\Big)^{-1}\int dE_ld\Omega_ld\Omega_p
       \frac{d\sigma(e^+e^-\to bl\nu\bar bW^-)}{d\rmp\;d\Omega_p\;d\Omega_l
         \;dE_l}(nl),
\end{equation}
where $l$ is the lepton four-momentum and $n$ may be any convenient
four-vector, for example (in the $t\bar t$ rest frame)
\begin{eqnarray}
  n_{_\parallel} = (0,{\bf n}_{_\parallel}) & & n_\perp = (0,{\bf n}_\perp)\\
  n_{\rm _N} = (0,{\bf n}_{\rm _N} & & n_{(0)} = (1,0,0,0).
\end{eqnarray}

This method has several advantages \cite{HJKP96}:
\begin{itemize}
\item it always leads to well-defined, physical, measurable quantities,
\item it is far easier than calculating the full multi-differential
  cross-section because the leptonic phase space can be integrated out
  completely, even in one of the first steps,
\item it still gives us the same information on {\bf P}, or more exactly,
  it can be used to define a quantity which is equivalent to {\bf P} in
  the absence of rescattering. Neglecting QCD corrections one easily derives
  from (\ref{fact})
  \begin{equation}\label{mom:born}
    \langle nl\rangle_{\rm Born} = \mbox{BR}(t\to bl\nu)\cdot\frac{1+2y+3y^2}{
      4(1+2y)}\cdot\Big[ (n\cdot t)+\frac{m_t}{3}(n\cdot P)\Big]
  \end{equation}
  where $t$ and $P$ are four-vectors given by
  \begin{equation}
    t = (m_t,\bfp) \qquad,\qquad P=(\frac{\rmp}{m_t}C_{_\parallel}^0\cos
        \vartheta,{\bf P}).
  \end{equation}
  The appearance of the 0-component of $P$ is welcome, because if $P$ is to be
  interpreted as a polarization vector, it has to fulfill the transversality
  condition $t\cdot P=0$, which indeed it does. Remember that we are working
  in the $t\bar t$ rest frame instead of the $t$ rest frame (the corresponding
  change in {\bf P} is of the order $\beta^2$ and thus does not appear here).
\end{itemize}

Rescattering modifies equation (\ref{mom:born}), the corrected expression reads
\begin{eqnarray}
  \langle nl\rangle &=& \mbox{BR}(t\to bl\nu)\cdot\frac{1+2y+3y^2}{4(1+2y)}
   \nonumber \\ && \cdot\Big[ (n\cdot t)(1+\varepsilon)+\frac{m_t}{3}(
   n\cdot P+n\cdot\delta P)\Big]
\end{eqnarray}
with
\begin{eqnarray}
 \varepsilon & = & \frac{y(1-y)}{(1+2y)(1+2y+3y^2)}C_{_\parallel}^0
    \psi_{\rm R}(\rmp,E)\cos\vartheta \\
 \delta {\rm P}_{_\parallel} &=& 
    \Bigg[\frac{2+3y-5y^2-12y^3}{(1+2y)(1+2y+3y^2)}-
    \frac{1-2y}{1+2y}(C_{_\parallel}^0)^2\Bigg]
      \psi_{\rm _R}(\rmp,E)\cos\vartheta \nonumber \\
    && +\frac{1-4y+3y^2}{4(1+2y+3y^2)}(1-3\cos^2\vartheta)C_{_\parallel}^0
    \psi_3(\rmp,E) \label{p2long}\\
 \delta {\rm P}_\perp &=& \frac{3(1-3y^2)}{2(1+2y+3y^2)}\psi_{\rm _R}(\rmp,E)
    \sin\vartheta \nonumber \\ 
    && -\frac{3(1-4y+3y^2)}{8(1+2y+3y^2)}C_{_\parallel}^0
    \psi_3(\rmp,E)\sin2\vartheta  \label{p2perp}\\
 \delta {\rm P}_{\rm _N} &=& 0,\qquad \delta{\rm P}_0 = 0.
\end{eqnarray}
The new function $\psi_3$ is defined as
\begin{eqnarray}
  \psi_3(\rmp,E) &=& \mbox{Im}\int\!\!\frac{d^3k}{(2\pi)^3}V(|\bfk-\bfp|)
   \frac{G(\rmk,E)}{G(\rmp,E)}\frac{\arctan\frac{|\bfk-\bfp|}{\Gamma_t}}
    {|\bfk-\bfp|}\nonumber \\ && \times \Bigg[3\Big(\frac{{\bf p
        \cdot(k-p)}}{\rmp|\bfk-\bfp|}\Big)^2-1\Bigg]
\end{eqnarray}
and its presence destroys the otherwise nice similarity between $P$ and
$\delta P$. Numerically its contribution is however negligible for realistic
top quark masses. The same applies to $\varepsilon$ so that at least
in practice the form of the Born-level result is recovered. One should
note that there was no a priori reason to expect this form to remain the
same.

\begin{figure}[tb]
\begin{center}\begin{tabular}{cc}
  \epsfxsize 65mm \mbox{\epsffile[0 0 567 454]{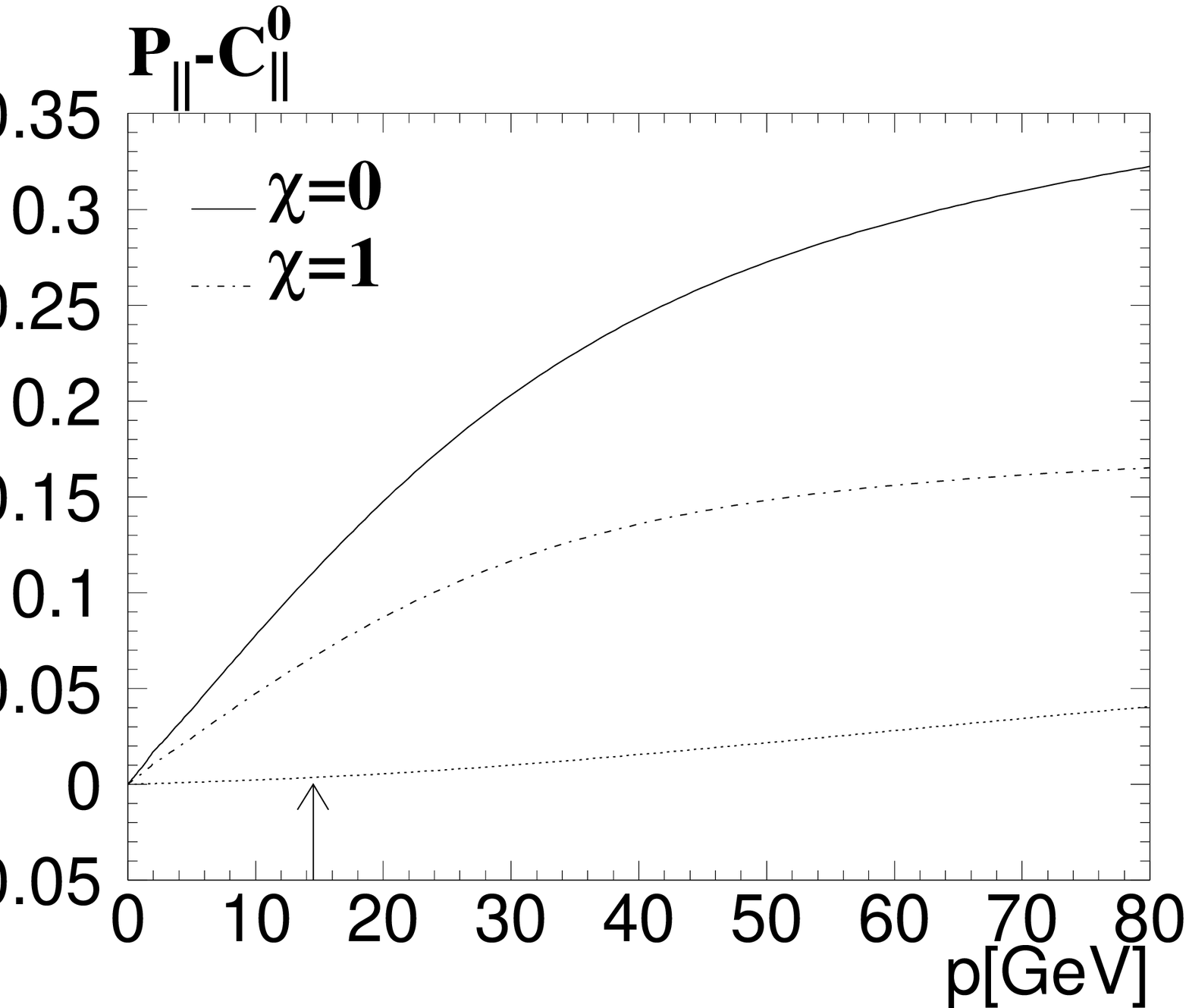}} &
  \epsfxsize 65mm \mbox{\epsffile[0 0 567 454]{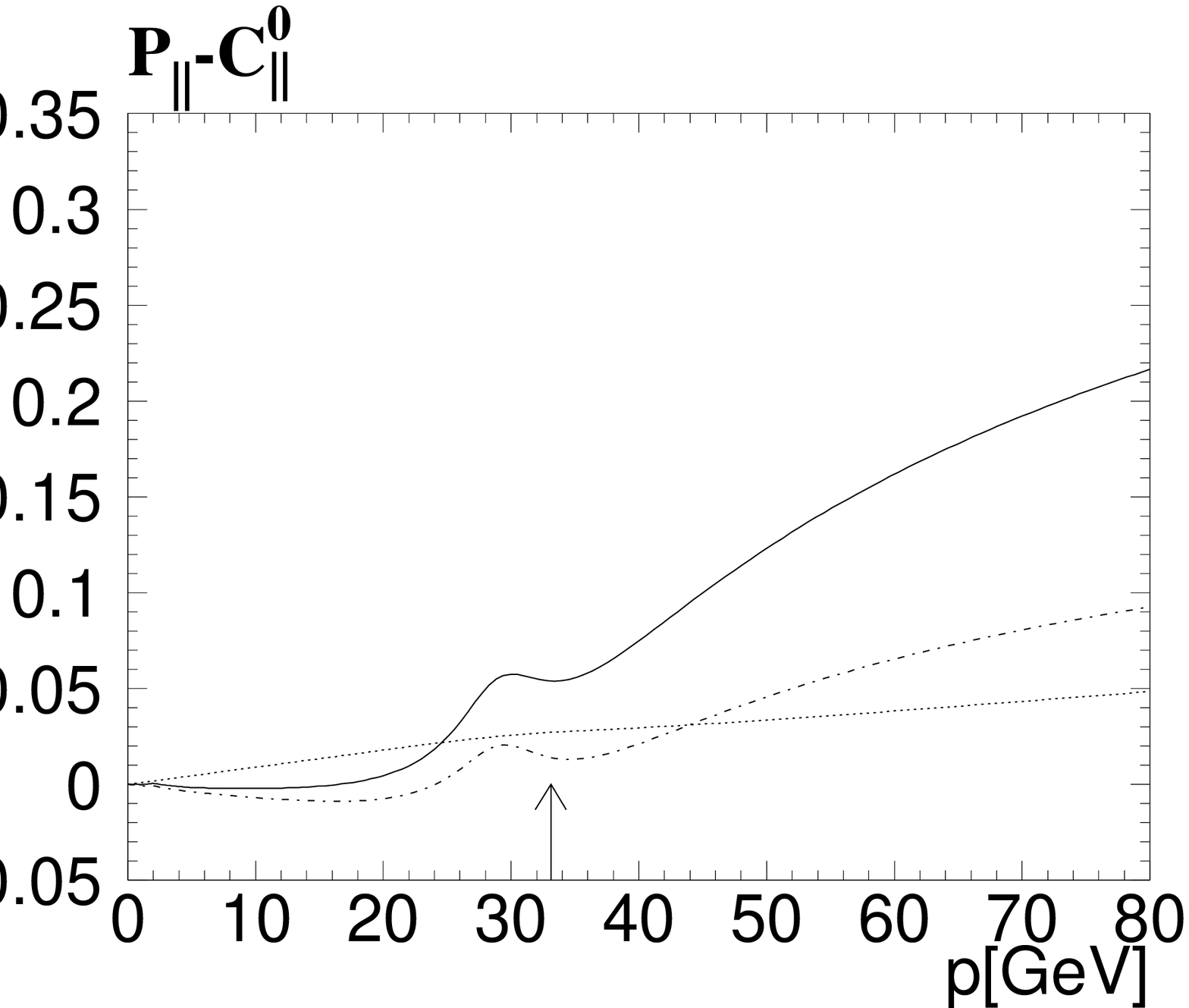}} \\
  a) E=-3\GeV & b) E=5\GeV
\end{tabular}\end{center}
\caption{\label{plong}Subleading part of the longitudinal component of the
  polarization vector P$_{_\parallel}$ for $\vartheta=0$.  The solid line
  shows the complete result for unpolarized beams, the dotted line the
  $S$-$P$--interference contribution. The dash-dotted line shows the complete
  result for fully polarized beams, in which case the $S$-$P$--wave
  contribution vanishes. The arrows indicate the position of the peak in the
  momentum distribution.}
\end{figure}

\begin{figure}[htb]
\begin{center}\begin{tabular}{cc}
  \epsfxsize 65mm \mbox{\epsffile[0 0 567 454]{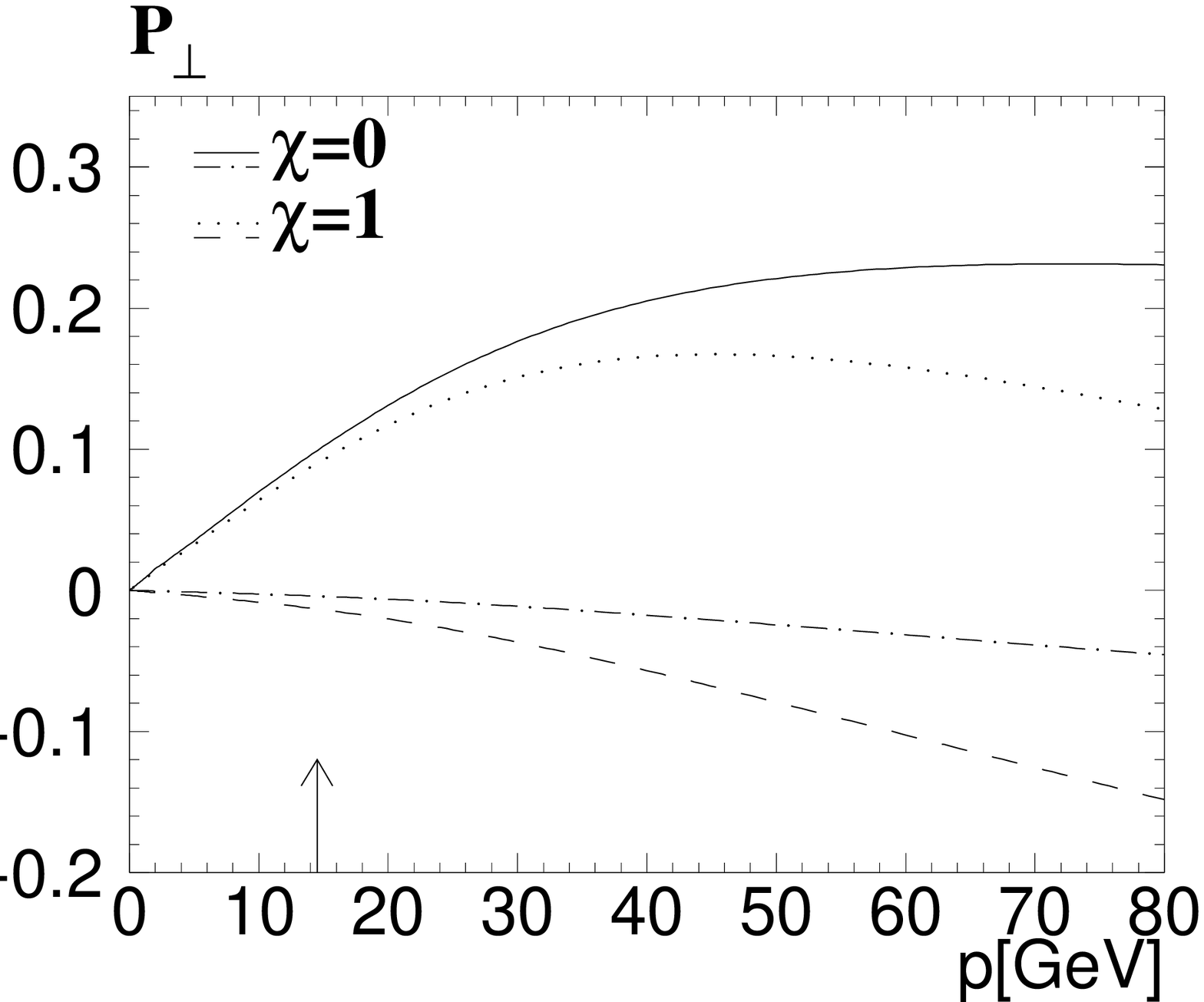}} &
  \epsfxsize 65mm \mbox{\epsffile[0 0 567 454]{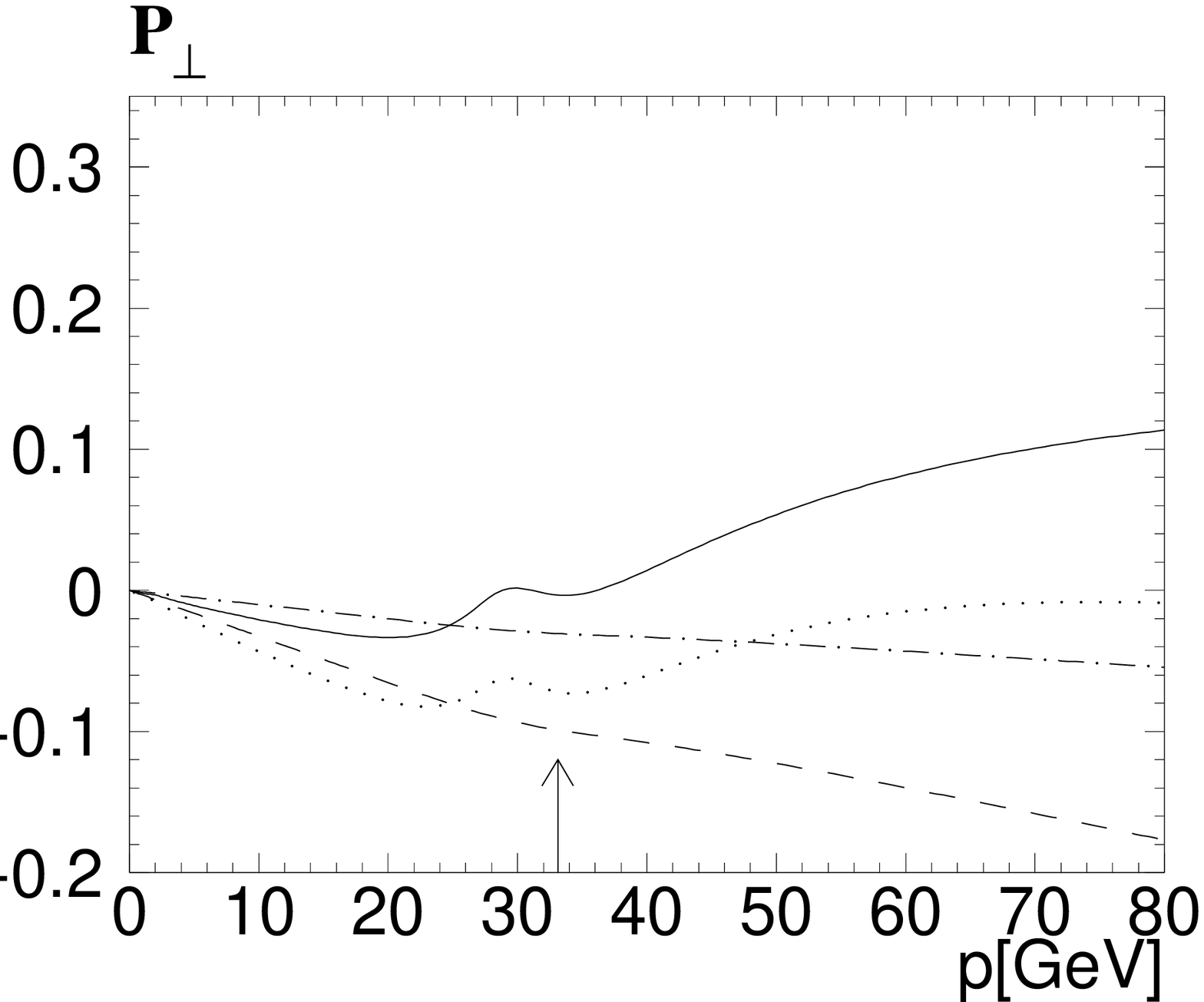}} \\
  a) E=-3\GeV & b) E=5\GeV
\end{tabular}\end{center}
\caption{\label{pperp}Transverse component of the polarization vector
  P$_\perp$ for $\vartheta=\pi/2$. The solid and dotted lines correspond to
  unpolarized beams, the dash-dotted and dotted lines to fully polarized
  beams. The solid and dash-dotted curves show the full result, the dashed and
  dotted the $S$-$P$--interference contribution. The arrows indicate the
  position of the peak in the momentum distribution.}
\end{figure}

In addition I should mention that $\psi_1$, $\psi_2$ and $\psi_3$ could be
interpreted as additional $S$-, $P$- and $D$-wave contributions, a viewpoint
motiviated by the way they are defined and the way they enter the expressions.

Figures \ref{plong} and \ref{pperp} show the predictions for the subleading
term of the longitudinal polarization and for the transverse part,
respectively. The solid and dash-dotted lines correspond to the full result
for unpolarized/fully polarized beams, the dashed and dotted lines to the
${\cal O}(\beta)$ term only. It is quite evident that rescattering has a
strong influence on the result, which is due to the fact that the ${\cal O}(
\beta)$-terms are suppressed by ratios of electroweak couplings appearing in
the coefficients $C_{_\parallel}^1$ and $C_\perp$. $\varphi_{\rm _R}$
and $\psi_{\rm _R}$ are of comparable size as they should be because
$\beta\sim\alpha_{\rm s}$.

%
%
\section{Summary and conclusions}

In this paper I have discussed recent results on observables in top quark
pair production near threshold at $e^+e^-$ colliders. I have sketched the
calculational procedure used to derive the predictions for the differential
cross-section and top quark polarization for arbitrary beam polarization
which are correct up to terms of the order $\rmp^2/m_t^2$ because they also
include the rescattering corrections. The method of moments proved to be
a very useful tool, which could of course also be used for other production
mechanisms or far above threshold.

It turned out that rescattering corrections are not generically small
and are indeed required for a consistent prediction of the components of
the top quark polarization vector lying in the production plane. They are
of minor importance for the forward-backward asymmetry but might
nevertheless be visible in precision studies, especially if polarized beams
are used. An important result was that there are no corrections
to the normal polarization. This component is of special interest because
it is the place where non-standard CP violating interactions would manifest
themselves. In connection with the leading term of P$_{_\parallel}$, which
can be measured by averaging over the top direction, i.e.\ via $\langle\langle
n_{_\parallel}\cdot P\rangle\rangle$, it provides an opportunity to
determine the top quark couplings $a_1\ldots a_4$.

I would like to express my gratitude to the organizers of the XXXVI Cracow
School of Theoretical Physics, especially M. Je\.zabek, for their invitation
and for providing such a nice atmosphere. I would also like to thank J.H.
K\"uhn for carefully reading the manuscript and giving useful advice.
%
%
\def\npb#1#2#3{{\it Nucl.~Phys.~}{\bf B #1} (#2) #3}
\def\plb#1#2#3{{\it Phys.~Lett.~}{\bf B #1} (#2) #3}
\def\prd#1#2#3{{\it Phys.~Rev.~}{\bf D #1} (#2) #3}
\def\pR#1#2#3{{\it Phys.~Rev.~}{\bf #1} (#2) #3}
\def\prc#1#2#3{{\it Phys.~Reports }{\bf C#1} (#2) #3}
\def\pr#1#2#3{{\it Phys.~Reports }{\bf #1} (#2) #3}
\def\prl#1#2#3{{\it Phys.~Rev.~Lett.~}{\bf #1} (#2) #3}
\def\sovnp#1#2#3{{\it Sov.~J.~Nucl.~Phys.~}{\bf #1} (#2) #3}
\def\yadfiz#1#2#3{{\it Yad.~Fiz.~}{\bf #1} (#2) #3}
\def\jetp#1#2#3{{\it JETP~Lett.~}{\bf #1} (#2) #3}
\def\zpc#1#2#3{{\it Z.~Phys.~}{\bf C #1} (#2) #3}

\end{document}